\documentclass[english,12pt]{article}
\usepackage{array}
\usepackage{graphicx}
\usepackage{amssymb}
\usepackage{amsmath}
\usepackage{multirow}
\usepackage{prettyref}
\usepackage{babel}
\usepackage{units}
\usepackage[latin1]{inputenc}
\usepackage{amsfonts}
\usepackage{amssymb}
\usepackage{babel}
\usepackage{cite}
\def\@fmsl@sh#1#2#3{\m@th\ooalign{$\hfil#1\mkern#2/\hfil$\crcr$#1#3$}}
 \def\eq#1\en{\begin{equation}#1\end{equation}}
\def\s[#1,#2]{[#1\stackrel{\star}{,}#2]}
\def\sx[#1,#2]{[#1\stackrel{\star_{x}}{,}#2]}

\newcommand{\nc}{\newcommand}
\nc{\beq}{\begin{equation}}
\nc{\eeq}{\end{equation}}
\nc{\beqa}{\begin{eqnarray}}
\nc{\eeqa}{\end{eqnarray}}

\def\bc{\begin{center}}
\def\ec{\end{center}}

\def\to{\rightarrow}

\def\gsim{\mathrel{\mathpalette\atversim>}}

\def\bc{\begin{center}}
\def\ec{\end{center}}

\def\gsim{\mathrel{\rlap{\lower4pt\hbox{\hskip1pt$\sim$}}

    \raise1pt\hbox{$>$}}}       

\def\gsim{\mathrel{\rlap{\lower4pt\hbox{\hskip1pt$\sim$}}
    \raise1pt\hbox{$>$}}}       

\begin{document}
\makeatletter
\def\fmslash{\@ifnextchar[{\fmsl@sh}{\fmsl@sh[0mu]}}
\def\fmsl@sh[#1]#2{%
  \mathchoice
    {\@fmsl@sh\displaystyle{#1}{#2}}%
    {\@fmsl@sh\textstyle{#1}{#2}}%
    {\@fmsl@sh\scriptstyle{#1}{#2}}%
    {\@fmsl@sh\scriptscriptstyle{#1}{#2}}}
\def\@fmsl@sh#1#2#3{\m@th\ooalign{$\hfil#1\mkern#2/\hfil$\crcr$#1#3$}}
\makeatother

\thispagestyle{empty}
\begin{titlepage}
\boldmath
\begin{center}
  \Large {\bf Self-healing of unitarity in Higgs inflation}
    \end{center}
\unboldmath
\vspace{0.2cm}
\begin{center}
{  {\large Xavier Calmet}$^{a,}$\footnote{x.calmet@sussex.ac.uk}
and {\large Roberto Casadio}$^{b,c,}$\footnote{roberto.casadio@bo.infn.it} }
 \end{center}
\begin{center}
{\sl $^a$Physics $\&$ Astronomy, 
University of Sussex,   Falmer, Brighton, BN1 9QH, United Kingdom 
}
\\
{\sl  $^b$Dipartimento di Fisica e Astronomia, Universit\`a di Bologna, via Irnerio 46, 40126 Bologna, Italy 
\\
 $^c$I.N.F.N., Sezione di Bologna, viale Berti Pichat 6/2, 40127 Bologna, Italy}
\end{center}
\vspace{5cm}
\begin{abstract}
\noindent
We reconsider perturbative unitarity violation in the standard model Higgs inflation model.
We show that the Cutkosky cutting rule implied by perturbative unitarity is fulfilled at one-loop. This is a strong indication that unitarity is restored order by order in perturbation theory. We then resum certain one-loop diagrams and show that the relevant dressed amplitude fulfills the Cutkosky rule exactly.
This is an example of the self-healing mechanism.  The original Higgs inflation model is thus consistent and does not require any new physics beyond the standard model at least up to the Planck scale.
\end{abstract}  
\end{titlepage}



\newpage

The Higgs boson of the standard model is often viewed as the source of a fine-tuning problem.
Indeed, the large hierarchy between the Planck scale, which sets the strength of the gravitational interaction,
and the weak scale, which is fixed by the parameters of the Higgs potential, has been the main motivation
to search for physics beyond the standard model.
The lack of experimental data to support physics beyond the standard model at the Large Hadron Collider
(LHC) could be seen as the second nail in the coffin for the naturalness problem.
Indeed, the very same argument applied to the cosmological constant would imply new physics at some
ridiculously low energy scale of the order of $10^{-3}\,$eV.
As in the case of the Higgs boson, there is no sign of new physics in cosmology associated with this energy scale.

Instead of being a source of problems, the Higgs boson might provide a solution to another type of
fine-tuning issue, namely that of the initial conditions of our universe.
The fine-tuning problematic at the beginning of our universe is very different from the fine-tuning problem
in the standard model.
One could argue that the naturalness issue of the standard model is in the eye of the beholder.  If one takes seriously the old-fashion pre-Wilsonian approach to  renormalization, quadratic divergences are not different from logarithmic ones: they are absorbed in the bare parameters during the renormalization process.  The Higgs mass is not calculable from first principles (see e.g. \cite{Iliopoulos:2006rc}). One should not have a theoretical expectation for the order of magnitude of its mass, it is a free parameter which needs to be measured in experiments.  From that perspective, the fine tuning issue of the Higgs mass is physically meaningless.

On the other hand, the fine-tuning issue in cosmology is really an initial condition problem.
Why did our universe start from such very specific initial conditions?
Inflation provides a natural framework to address this question.
It has been shown in Ref.~\cite{Bezrukov:2007ep,Barvinsky:2008ia,DeSimone:2008ei,Bezrukov:2013fka} that the Higgs boson with a fairly large non-minimal
coupling to the Ricci scalar could play the role of the inflaton. 

This large non-minimal coupling is the source of a potential issue with perturbative unitarity
(see, e.g.~\cite{Lerner:2009na,Burgess:2010zq,Atkins:2010yg,Hertzberg:2010dc} and references therein).
As we shall see shortly, unitarity is believed to be violated at an energy scale of $M_P/\xi$ in
today's Higgs vacuum, while it would be violated at a scale  $M_P/\sqrt{\xi}$ in the inflationary
background.
The breakdown of perturbative unitarity is a sign of strong dynamics or new physics which kicks
in at the scale of the breakdown of perturbative unitarity, thereby restoring unitarity.
However, both new physics and strong dynamics could jeopardize the flatness of the scalar potential
which is needed to obtain the correct number of e-folding required to explain the flatness of our universe.
 
In this work, we reconsider the issue of the breakdown of perturbative unitarity in the Higgs inflation
model (see, e.g.~\cite{Atkins:2010yg} and references therein).
Let us start with the Higgs boson doublet $H$, non-minimally coupled to the Ricci scalar $ \mathcal{R}$,
as described by the action
\begin{eqnarray}
\label{action2}
S
=
\int d^4x \, \sqrt{-g} \left[ \left( \frac12  M^2
+ \xi H^\dagger H \right) \mathcal{R}
- (D^\mu H)^\dagger (D_\mu H)
+ \mathcal{L}_{SM}  \right]
\ .
\end{eqnarray}
We will take the Higgs doublet of the form $H=1/\sqrt{2}\left(\pi^+, \bar \phi +\phi+ i \pi^0\right)^\top$,
where $\bar \phi$ is some background value for the physical Higgs boson,
$\pi^i$ are the pseudo-Goldstone bosons and $\phi$ will lead to the physical Higgs boson.
In today's universe, the background value of the Higgs boson would be the vacuum expectation
value $v=246$ GeV.
Successful inflation requires a fairly large non-minimal coupling $\xi\sim 10^4$~\cite{Bezrukov:2007ep,Barvinsky:2008ia}. Interestingly, it was recently pointed out that radiative corrections are large and that a smaller coupling might suffice~\cite{Allison:2013uaa}. Note that the LHC data allows one to bound the Higgs boson's non-minimal coupling. The current data implies that $\xi>2.6 \times 10^{15}$ is excluded at the $95\%$ confidence
level~\cite{Atkins:2012yn}.

It is useful to define a background dependent Planck mass given by
\begin{eqnarray}
\label{effPlanck}
M^2+\xi \bar \phi^2
=
M_P(\bar \phi)^2
\ ,
\end{eqnarray}
and linearize the gravitational metric using
\begin{eqnarray}
g_{\mu\nu}=\bar g_{\mu\nu} + h_{\mu\nu} 
\ ,
 \end{eqnarray}
where $\bar g_{\mu\nu}$ is a background metric and $h_{\mu\nu}$ is the graviton.
Using the linearized fields, one obtains the kinetic terms for the graviton, the Higgs boson and its
pseudo-Goldstone bosons:
\begin{eqnarray}
L^{(2)}
&=&
-\frac{ M_P(\bar \phi)^2}{8}
\left(h^{\mu\nu}  \Box h_{\mu\nu} +
2  \partial_\nu h^{\mu\nu}  \partial^\rho h_{\mu\rho} - 2  \partial_\nu h^{\mu\nu}  \partial_\mu h - h\Box h
\right)
\nonumber
\\
&&
+ \frac{1}{2} \left(\partial_\mu \phi\right)\left(\partial^\mu \phi\right)
+\frac{1}{2} \left(\partial_\mu \pi^0\right) \left(\partial^\mu \pi^0\right)
+\left(\partial_\mu \pi^+\right)\left(\partial^\mu \pi^-\right)
\nonumber
\\
&&
+\xi \bar \phi\left( \Box  h - \partial_\lambda  \partial_\rho h^{\lambda\rho}\right)\phi
\ ,
 \end{eqnarray}
where $\Box=\partial_\alpha\partial^\alpha$.
The pseudo-Goldstone bosons are canonically normalized, but there is a mixing between
the kinetic terms of the graviton and that of the Higgs boson.
Taking
\begin{eqnarray}
 \phi
 &=&
\frac{1}{\sqrt{1 + \frac{6 \xi^2 \bar \phi^2}{M_P^2(\bar \phi)}}} \hat \phi
\\ 
 h_{\mu\nu}
 &=&
 \frac{1}{M_P(\bar \phi) } \hat h_{\mu\nu}
 -\frac{2 \xi \bar \phi}{M^2_P(\bar \phi)\sqrt{1+\frac{6 \xi^2 \bar \phi^2}{M_P^2(\bar \phi)}}}
 \bar g_{\mu\nu} \hat \phi
 \end{eqnarray}
leads to a rescaling of the couplings of the Higgs boson to all particles of the standard model
and to the linearized Ricci scalar.
The latter appears in the term
 \begin{eqnarray}
 \frac{\xi}{ M_P(\bar \phi) \left (1+ \frac{6 \xi^2 \bar \phi^2}{M_P^2(\bar \phi)} \right )} \hat \phi^2 \Box \hat h. 
  \end{eqnarray}
Note also that the non-minimal coupling constant of the pseudo-Goldstone bosons to the linearized Ricci scalar
is not affected by the rescaling of the Higgs boson's wavefunction:
 \begin{eqnarray}
 \frac{\xi}{ M_P(\bar \phi)}
 \left[\pi^+ \pi^- +(\pi^0)^2\right] \Box \hat h
 \ . 
  \end{eqnarray}
  
In Ref.~\cite{Bezrukov:2010jz}, the coefficient of $\hat \phi^2 \Box \hat h$ is identified as the cutoff
of the effective theory. 
This cutoff is then shown to behave i)~as $M/\xi$ for small Higgs boson background field values ($\bar \phi \ll M/\xi$),
ii)~as $\xi \bar \phi^2/M$ in the intermediate region ($ M/\xi \ll \bar \phi \ll M/\sqrt{\xi}$), and iii)~as $\sqrt{\xi} \bar \phi$
for large background field values ($\bar \phi\gg M/\sqrt{\xi}$).
  
While the energy scale suppressing the higher dimensional operator $\hat \phi^2 \Box \hat h$ provides
some feeling of the scale at which the effective theory may break down, it is difficult to make precise
statements solely based on dimensional analysis.
We will instead reconsider the perturbative unitarity bound in the light of the recent paper by Aydemir, Anber
and Donoghue~\cite{Aydemir:2012nz}.
In this remarkable paper, they show that the one-loop correction to the gravitational elastic scattering of scalars
is sufficient to unitarize the tree-level amplitude, which grows with the center-of-mass energy squared.
In particular, they show that $|T_2^{\rm tree}|^2={\rm Im}\,(T_2^{\rm 1-loop})$ for the $J=2$ partial-wave.

We here extend their result to the $J=0$ partial-wave, which is the relevant one for obtaining some sensitivity
on the non-minimal coupling of the Higgs.
We cannot directly apply their results, since the non-minimal coupling constant for the rescaled
Higgs field wavefunction $\hat \phi$ is, in general, different from that of the pseudo-Goldstone bosons $\pi^i$.
However, for $\bar \phi= v$ (the present vacuum expectation value of the Higgs field), we find
  \begin{eqnarray}
 \frac{\xi}{M_P(\bar \phi)  \left (1+ \frac{6 \xi^2 v^2}{M_P^2}\right)} \hat \phi^2 \Box \hat h
 =
 \frac{\xi}{M_P} \hat \phi^2 \Box \hat h 
 + {\cal O}\left(\frac{1}{M_P^3}\right)
 \ ,
  \end{eqnarray}  
with $M_P^2=M^2+\xi v^2$, so that  we can safely neglect the rescaling as long as we consider tree level
and one-loop amplitudes:
in that approximation, the Higgs boson and the pseudo-Goldstone bosons couple to the Ricci scalar
with the same non-minimal coupling.
 
But,  for large background field values, the Higgs field and the pseudo-Goldstone bosons do not have the same non-minimal coupling.
In fact, in this regime we find
\begin{eqnarray}
 \frac{\xi}{ M_P(\bar \phi) \left(1+ \frac{6 \xi^2 \bar \phi^2}{M_P^2(\bar \phi)}\right)} \hat \phi^2 \Box \hat h
 \to
 \frac{1}{6 \sqrt{\xi} \bar \phi} \hat \phi^2 \Box \hat h 
  \end{eqnarray}
 for the Higgs boson,
 and 
 \begin{eqnarray}
 \frac{\xi}{ M_P(\bar \phi)}  \left[\pi^+ \pi^- +(\pi^0)^2\right] \Box \hat h
 \to
 \frac{\sqrt{\xi} }{\bar \phi} \left[\pi^+ \pi^- +(\pi^0)^2\right] \Box \hat h 
 \end{eqnarray}
for the pseudo-Goldstone bosons.
We thus have to generalize the calculations performed in
Ref.~\cite{Aydemir:2012nz} to the case of scalar fields with different non-minimal couplings.

The tree level amplitude for the gravitational elastic scatting of the two Higgs bosons (with a non-minimal coupling $\xi_1$) into two other scalars (with a non-minimal coupling $\xi_2$)
is given by
\begin{eqnarray}
A_{\rm tree}
=\frac{8 \pi G_N(\bar \phi)}{s} \left [ s^2 \left (6 \xi_1 \xi_2 + \xi_1 + \xi_2  \right ) + u t \right].
\end{eqnarray}
The background dependent Newton's constant is given by
\begin{eqnarray}
G_N(\bar \phi)
=
\frac{1}{8 \pi (M^2+\xi \bar \phi^2)}
\ .
\end{eqnarray} 
Clearly for $\xi_1=\xi_2$ one recovers the result of ~\cite{Aydemir:2012nz}. 

We now consider the one-loop amplitude with scalar fields which modify the propagator of the graviton. It is important to realize that the non-minimal coupling on both vertices in the loop must be the same. One can thus use directly the expression (which we have verified) given in ~\cite{Aydemir:2012nz} for the one-loop quantum corrected graviton propagator given by
 \begin{eqnarray}
 i D^{\alpha \beta \mu \nu}_{1-loop}= \frac{i}{2 q^2} (1+ 2 F_2(q^2)) [L^{\alpha \mu} L^{\beta \nu} +L^{\alpha \nu} L^{\beta \mu} -L^{\alpha \beta} L^{\mu \nu}] -i \frac{F_1(q^2)}{4} L^{\alpha \beta}L^{\mu\nu},
 \end{eqnarray} 
where $L^{\alpha \beta}=\eta^{\alpha\beta}-q^\alpha q^\beta/q^2$ and
\begin{eqnarray}
F_1(q^2)=-\frac{1}{30 \pi} N_s G_N(\bar \phi)(1+10\xi + 30 \xi^2) \log\left( \frac{-q^2}{\mu^2}\right)
\end{eqnarray} 
and
\begin{eqnarray}
F_2(q^2)=\frac{1}{240}N_s G_N(\bar \phi) q^2  \log\left( \frac{-q^2}{\mu^2}\right)
\end{eqnarray} 
where $N_s$ is the number of scalar fields in the loop and $\mu$ is the renormalization scale associated with the operators $R^2$ and $R_{\mu\nu}R^{\mu\nu}$.

The one-loop amplitude for $\phi_A + \phi_A \to  \phi_B+ \phi_B$ with an arbitrary scalar field in the loop with a non-minimal coupling $\xi$ is given by 
\begin{eqnarray}
A_{\rm 1-loop}(\xi_A,\xi_B,\xi)
=
-\frac{G^2_N(\bar \phi) }{15} \left[ s^2 F_3(\xi_A,\xi_B,\xi)- u t \right] \log(-s)
\ ,
\end{eqnarray}
where 
\begin{eqnarray}
F_3(\xi_A,\xi_B,\xi)&=& 1+10 \xi + 5 \xi_A + 5 \xi_B+ 30 \xi^2  +60 \xi \xi_A + 60 \xi \xi_B + 30 \xi_A \xi_B
\\ \nonumber &&
+180 \xi^2 \xi_A +180 \xi^2 \xi_B + 360  \xi \xi_A \xi_B +1080 \xi^2 \xi_A \xi_B. 
\end{eqnarray}
The nonminimal coupling of the scalar field in the loop is denoted by $\xi$, that of the initial state scalars by $\xi_A$ and that of the final state scalars by $\xi_B$.  Again in the limit $\xi_i \to \xi$, we recover the result of~\cite{Aydemir:2012nz}. 
Note that we  are considering real scalar fields here, as one can rewrite the pseudo-Goldstone bosons $\pi^\pm$ in terms of real fields.

For the Higgs boson of the standard model we find:
\begin{eqnarray}
A_{\rm 1-loop}(\phi + \phi \to \pi^i + \pi^i)= A_{\rm 1-loop}(\xi_{H},\xi_{G},\xi_{H}) + 3 A_{\rm 1-loop}(\xi_{H},\xi_{G},\xi_{G}) 
\end{eqnarray}
where $\xi_{H}$ is the nonminimal coupling of the Higgs boson to the Ricci scalar and $\xi_{G}$ is that of the pseudo-Goldstone bosons.

We can then derive the $J=0$ partial-wave amplitudes using the standard procedure, namely
\begin{eqnarray}
a_l(s)=\frac{1}{32 \pi} \int^1_{-1} \mbox{d} \cos(\theta) P_l( \cos(\theta) ) A(s,\theta)
\end{eqnarray} 
where $P_l$ are the Legendre polynomials. We calculate the partial-waves for  the transitions $\phi+\phi \to \pi^i +\pi^i$, $\pi^i+\pi^i \to \pi^j +\pi^j$, $\phi+\phi \to \phi+\phi $ and $ \pi^i + \pi^i \to \phi+\phi$ at tree level. To be precise, we consider the elastic scattering between initial/final states $1/\sqrt{2}  | \phi \phi \rangle$  and $1/\sqrt{2}| \pi^i \pi^i \rangle$.  We find a $4 \times 4$ partial-wave matrix.
\begin{eqnarray}
a_{0, {\rm tree}}
&=&
\frac{G_N(\bar \phi) \,s }{12} \\ \nonumber && \times
\left (\begin{array}{cccc}
(6 \xi_{H}+1)^2 & (6 \xi_{H}+1) (6 \xi_{G}+1)& (6 \xi_{H}+1) (6 \xi_{G}+1) &(6 \xi_{H}+1) (6 \xi_{G}+1) \\
(6 \xi_{H}+1) (6 \xi_{G}+1) & (6 \xi_{G}+1)^2&  (6 \xi_{G}+1)^2 &(6 \xi_{G}+1)^2 \\
(6 \xi_{H}+1) (6 \xi_{G}+1) & (6 \xi_{G}+1)^2& (6 \xi_{G}+1)^2 &(6 \xi_{G}+1)^2 \\
(6 \xi_{H}+1) (6 \xi_{G}+1) &  (6 \xi_{G}+1)^2&  (6 \xi_{G}+1)^2 &(6 \xi_{G}+1) ^2\\
\end{array} \right).
\end{eqnarray} 
The only non-zero eigenvalue of the matrix is 
\begin{eqnarray}
a_{0, {\rm tree,max}}
&=&
\frac{G_N(\bar \phi) \,s }{3}\left (9 \xi_{H}^2 + 3 \xi_{H} + 27 \xi_{G}^2 + 9 \xi_{G} +1 \right) 
\end{eqnarray} 
while the non-zero partial-wave for the one-loop diagram is given by  
 \begin{eqnarray}
a_{0, {\rm 1-loop,max}}
&=& - \frac{G^2_N(\bar \phi)\, s^2 }{9 \pi} \left (9 \xi_{H}^2 + 3 \xi_{H} + 27 \xi_{G}^2 + 9 \xi_{G} +1 \right) ^2
 \log(-s)
\ ,
 \end{eqnarray} 
so that 
 \begin{eqnarray}
\mbox{Im}\left(  a_{0, {\rm 1-loop, max}}\right) &=& \frac{G^2_N(\bar \phi)\, s^2 }{9} \left (9 \xi_{H}^2 + 3 \xi_{H} + 27 \xi_{G}^2 + 9 \xi_{G} +1 \right) ^2.
 \end{eqnarray} 
It is straightforward to check that 
\begin{eqnarray}
|a_{0, {\rm tree,max}}|^2=\mbox{Im} \left( a_{0, {\rm 1-loop,max}}\right)
\ ,
\end{eqnarray} 
as required by perturbative unitarity.
This result holds for any background values of the Higgs and gravitational fields
and thus is valid in today's universe as it is at the time of inflation. 

Note that this is only a one-loop result. As observed for the $J=2$ partial-wave  in \cite{Aydemir:2012nz,Han:2004wt}, one can rewrite the $J=0$ partial-wave to one-loop order as
\begin{eqnarray} \label{a0}
a_0=a_0^{(1)} \left ( 1 + \frac{\mbox{Re} \ a_0^{(2)}}{a_0^{(1)}} + i a_0^{(1)} \right),
\end{eqnarray} 
where the superscript denotes the  order in $G_N(\phi) s$. This equation is derived using $|a_{0, {\rm tree}}|^2=\mbox{Im} \left( a_{0, {\rm 1-loop}}\right)$ which we have just shown and the fact that  $a_0^{(1)}$ is real. One notices that $a_0$ as given in Eq. (\ref{a0})  is the first term of the geometric series generated by the resummation of the one-loop diagrams. Resumming this series, one finds:
\begin{eqnarray}
a_0=\frac{a_0^{(1)}}{1 -  \mbox{Re} \ a_0^{(2)}/a_0^{(1)} - i a_0^{(1)}}.
\end{eqnarray} 
 The resummed amplitude satisfies exactly $|a_{0}|^2=\mbox{Im} \left( a_{0}\right)$. Unitarity is thus respected at arbitrary energies, despite the tree-level violation. We have used the same resummation technique as the one used in \cite{Aydemir:2012nz,Han:2004wt} namely a Pad\'e type resummation of the graviton propagator. 
 
 One can also verify the Cutkosky rule at the non-perturbative level by resumming  an infinite series of vacuum polarization diagrams.  We will work in the limit where the Higgs boson and the pseudo-goldstone bosons have the same non-minimal coupling to curvature. Also we will work in the limits $\xi\gg1$ and large $N$ limit where $N=4$ (one may wonder whether $N$ is large enough to use this approximation here, but note that, for example, the large N limit with $N=3$ works well in QCD). In the large $\xi$ limit, the one-loop quantum corrected propagator is given by 
\begin{eqnarray}
 i D^{\alpha \beta \mu \nu}_{1-loop}= -\frac{i}{2 s} \left (1+\frac{F_1(s)}{2} \right) L^{\alpha \beta} L^{\mu \nu}.
 \end{eqnarray} 
Resumming an infinite series of one-loop diagrams in the large $\xi$ and large N limits but keeping $\xi G_N N$ small, we find
 \begin{eqnarray}
 i D^{\alpha \beta \mu \nu}_{dressed}= -\frac{i}{2 s} \frac{L^{\alpha \beta} L^{\mu \nu}}{\left (1- \frac{ s F_1(s)}{2}\right) }.
 \end{eqnarray} 
 Note that $F_1(s)$ is negative, there is thus no pole in the propagator.  The dressed amplitude in the large $\xi$ and large $N$ limits is given by
  \begin{eqnarray}
  A_{dressed}=\frac{48 \pi G_N(\bar \phi) s  \xi^2}{1+\frac{2}{\pi} G_N(\bar \phi)  s \xi^2 \log(-s/\mu^2)} 
    \end{eqnarray}
One easily verifies that the $J=0$ partial-wave dressed amplitude fulfills 
\begin{eqnarray}
|a_{0}|^2=\mbox{Im} \left( a_{0}\right).
\end{eqnarray} 
In other words, unitarity is restored within general relativity without any new physics or strong dynamics (we are keeping $\xi G_N$ small). We wish to emphasize an important point. It was pointed out in \cite{Han:2004wt} (see also \cite{Tomboulis:1977jk}), that $1/N$ resummations 
can lead to amplitudes which are unitary, but where a pair of complex-conjugate poles on the physical sheet violate the usual analyticity properties. In our case,  as mentioned already, we do not have a pole in the propagator. It is worth mentioning that we are not only resumming bubble diagrams in the large $N$ limit but in the large $\xi$ limit and large $N$ limit. The large $N$ limit is important as it  allows one to justify for example that a diagram with two one loop bubble involving a scalar field on a graviton line is larger  by a factor of $N$ than a two loop diagram with one bubble involving a scalar field and with a graviton being exchanged in the scalar loop. While Pad\'e type resummations sometimes lead to pathologies in the form of tachyons \cite{Coleman:1974jh}, here we cannot identify any obvious problem with the resummation we perfomed. 

Our result  is  a strong indication that the self-healing mechanism is at work and that no new physics is needed below the background dependent Planck scale.  Note that the loop corrections are all small compared to the tree level amplitude and since the scale of inflation is always below the unitarity violation scale, one needs not to worry about the stability of the inflationary potential. There is no large correction, which could affect the flatness of the potential.  Finally, let us remark that physics is not frame dependent  \cite{Calmet:2012eq} and although we did our calculations in the Jordan frame, it would be straightforward to do them in the Einstein frame. 

{\it Conclusions:}
We have shown by resumming a certain class of one-loop diagrams that the Higgs boson can be the inflaton without the need for new physics beyond the standard model and general relativity. Since loop corrections are small,
one does not need to worry about the flatness of the potential as it does not receive sizeable corrections.
Our work validates the original minimalistic standard model Higgs inflation scenario.

{\it Acknowledgments:}
X.C. would like to thank  H.-J. He for helpful discussions. This work is supported in part by the European Cooperation in Science and Technology (COST) action MP0905 ``Black Holes in a Violent  Universe" and by the Science and Technology Facilities Council (grant number  ST/J000477/1).


\bigskip{}

\end{document}